\documentclass[cits]{PoS}

\title{Lithium-rich stars in the Sloan Digital Sky Survey}

\ShortTitle{Lithium-rich SDSS stars}

\author{\speaker{Sarah L. Martell}\\
        Australian Astronomical Observatory\\
        E-mail: \email{smartell@aao.gov.au}}

\author{Matthew D. Shetrone\\
        McDonald Observatory, University of Texas Austin\\
        E-mail: \email{shetrone@astro.as.utexas.edu}}

\abstract{We report the discovery of 23 lithium-rich post-main-sequence stars, identified from moderate-resolution SDSS spectroscopy and confirmed with high-resolution spectra taken at the Hobby-Eberly Telescope. These new Li-rich stars cover a broad range in mass and evolutionary phase, including bright giants and post-AGB stars. The process responsible for preserving or producing excess lithium in a small fraction of evolved stars remains unclear.}

\FullConference{XII International Symposium on Nuclei in the Cosmos\\
		 August 5-12, 2012\\
		 Cairns, Australia}

\begin{document}

\section{Introduction}
Lithium in the Universe is an intriguing subject of study, because there is less of it in stars than standard big bang nucleosynthesis predicts\cite{MB10}\cite{KG06}, while roughly $1\%$ of post-main-sequence stars have significantly more lithium than models of stellar evolution predict\cite{CB00}. It is produced in a sub-cycle of the proton-proton chain, but it is quickly destroyed at temperatures above $2.6 \times 10^{6}$ K because its cross section to proton capture is quite large.

The first Li-rich red giant was discovered serendipitously\cite{WS82}, and later studies showed that they are quite rare\cite{PS00}. Six independent studies in the past three years\cite{GZ09}\cite{KR11}\cite{RF11}\cite{MV11}\cite{LU12}\cite{KF12} have identified 47 new Li-rich post-main-sequence stars, in the Milky Way and in Local Group dwarf galaxies. These stars cover a broad range of parameters and abundances, from very low-metallicity RGB stars in Local Group dwarf galaxies to Solar-metallicity RGB stars in the Galactic bulge. This heterogeneity is a challenge for existing models of Li production, and it is unclear what process or processes are responsible.

\section{The observational data set}
We initially selected 8535 stars from the seventh data release\cite{DR7} of the Sloan Digital Sky Survey (SDSS DR7), requiring that the stars have gravity near or below the gravity of the ``bump'' in the RGB luminosity function, and that the spectra and the derived stellar parameters be relatively clean (signal-to-noise ratio per pixel above 40, $\sigma_{\rm [Fe/H]} \leq 0.3$, $\sigma_{\rm log(g)} \leq 0.5$). For each of those spectra, we measured the index $S(Li)$:

\begin{equation}
S(Li) = -2.5 \times \log \frac{\int_{6706}^{6713} I_{\lambda}d\lambda}{\int_{6730}^{6738} I_{\lambda}d\lambda}
\end{equation}

\begin{figure}
\includegraphics[width=.6\textwidth]{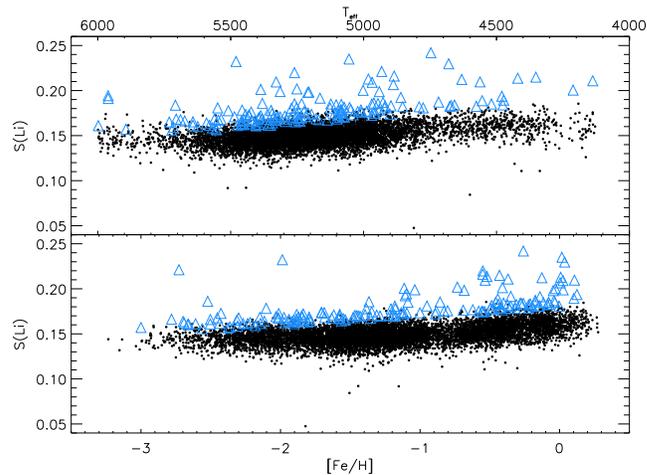}
\caption{Lithium line strength index $S(Li)$ versus $T_{eff}$ and [Fe/H] for our initial data set. Stars that are outliers in both distributions are shown as open blue triangles.}
\label{fig1}
\end{figure}

A larger index value indicates stronger absorption in the $6708\hbox{\AA}$ Li line. $S(Li)$ is a mild function of both effective temperature and metallicity, so we selected stars that are outliers in both parameters simultaneously as potential followup candidates. Figure \ref{fig1} shows the distributions of $T_{eff}$ and [Fe/H] for our initial data set, and the 162 stars with relatively strong $S(Li)$ are shown as open blue triangles.

We then selected 36 likely-looking candidates for high-resolution followup spectroscopy using the Hobby-Eberly Telescope at McDonald Observatory. For 23 of those, the spectra we obtained were of sufficient quality to allow us to determine effective temperatures, gravities, metallicities and lithium abundances. All 23 of these stars have log$\epsilon$(Li)$\ge 1.9$, with the majority above 2.5 and a few as high as 4.5. Figure 2 shows log$\epsilon$(Li) versus [Fe/H] for our 23 Li-rich stars, with lower-RGB stars shown as green squares, intermediate-age red clump stars shown as purple circles, old RGB stars shown as blue triangles and post-AGB stars shown as black crosses. There are no obvious trends between Li abundance and other stellar properties.

\begin{figure}
\includegraphics[width=.6\textwidth]{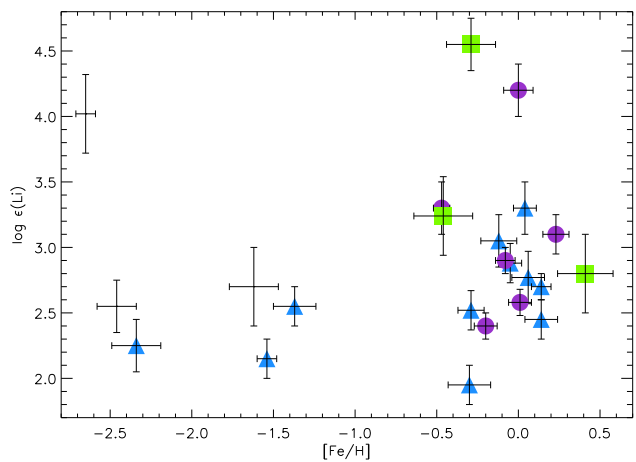}
\caption{Lithium abundance versus metallicity for our 23 Li-rich stars. Lower-RGB stars are shown as green squares, intermediate-age red clump stars as purple circles, old RGB stars as blue triangles and post-AGB stars as black crosses.}
\label{fig2}
\end{figure}

\section{Discussion}
Figure 3 shows the locations of our 23 Li-rich stars in the $T_{eff}$-log(g) plane, with representative isochrones overplotted. There are lower-RGB stars (green squares), intermediate-age red clump stars (purple circles), low-mass RGB stars (blue triangles) and post-AGB stars (black crosses).

\begin{figure}
\includegraphics[width=.6\textwidth]{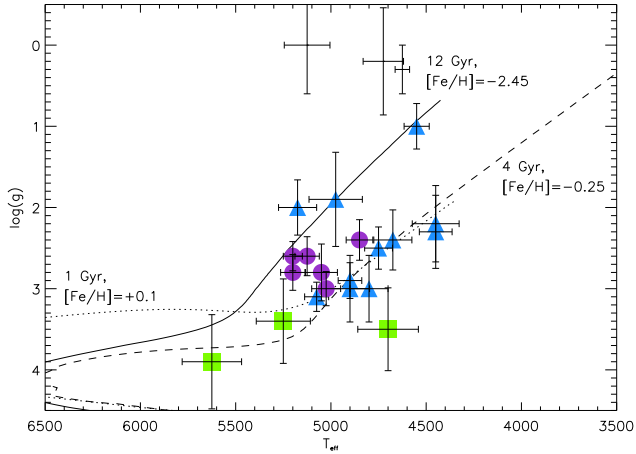}
\caption{Our 23 Li-rich stars in the $T_{eff}$-log(g) plane, with the same symbol and color coding as in Fig. 2.}
\label{fig3}
\end{figure}

It has been suggested\cite{CB00} that the onset of deep mixing\cite{MS08}\cite{AS12}, the point in RGB evolution when the H-burning shell encounters the $\mu$-discontinuity left behind at the end of first dredge-up, is related to a brief phase of lithium production. This happens at the RGB bump for low-mass stars, and at the red clump for intermediate-mass stars. Many of our stars are near the RGB bump or the red clump, but some are brighter giants and post-AGB stars, and their lithium abundances cannot be explained as an effect of the onset of deep mixing.

We agree with \cite{KF12} that Li-rich field giants are perfectly normal stars in most respects, but based on this data we must disagree with their conclusion that the onset of deep mixing produces a brief phase of Li enrichment in all stars. Such a universal process would imply that Li-rich stars should only be found near the RGB bump or the red clump, which they are not. We propose instead that deep mixing is capable of producing Li enrichment at any point above the RGB bump, based on thermohaline mixing models (A. Karakas, private communication). This does not explain what triggers a phase of Li production, but does fit comfortably with the the wide range of evolutionary phase that Li-rich stars occupy, and the short duration of the phase, which is inferred from the rarity of these stars.

\section{Acknowledgements}
SLM acknowledges the support of the Sonderforschungsbereich SFB 881 ``The Milky Way System'' (subprojects A2 and A5) of the German Research Foundation (DFG).

Funding for SDSS-III has been provided by the Alfred P. Sloan Foundation, the Participating Institutions, the National Science Foundation, and the U.S. Department of Energy. The SDSS-III web site is http://www.sdss3.org/. SDSS-III is managed by the Astrophysical Research Consortium for the Participating Institutions of the SDSS-III Collaboration including the University of Arizona, the Brazilian Participation Group, Brookhaven National Laboratory, University of Cambridge, University of Florida, the French Participation Group, the German Participation Group, the Instituto de Astrofisica de Canarias, the Michigan State/Notre Dame/JINA Participation Group, Johns Hopkins University, Lawrence Berkeley National Laboratory, Max Planck Institute for Astrophysics, New Mexico State University, New York University, Ohio State University, Pennsylvania State University, University of Portsmouth, Princeton University, the Spanish Participation Group, University of Tokyo, University of Utah, Vanderbilt University, University of Virginia, University of Washington, and Yale University.

The Hobby-Eberly Telescope (HET) is a joint project of the University of Texas at Austin, the Pennsylvania State University, Ludwig-Maximillians-Universit\"{a}t M\"{u}nchen, and Georg-August-Universit\"{a}t G\"{o}ttingen. The HET is named in honor of its principal benefactors, William P. Hobby and Robert E. Eberly.

\end{document}